\documentclass[preprintnumbers,nofootinbib,superscriptaddress]{revtex4}
\usepackage[dvipdfmx]{graphicx} 
\usepackage{amsmath,amssymb,amsfonts,bm}

\def\a{\alpha}  \def\g{\gamma}  \def\d{\delta} \def\D{\Delta} \def\e{\epsilon}   \def\th{\theta}   \def\l{\lambda}  \def\m{\mu} \def\n{\nu}      \def\s{\sigma}  \def\t{\tau} \def\ph{\phi}     \def\o{\omega} 

\def\dg{\dagger}  

\newcommand{\meV}{ {\rm meV} }  \newcommand{\keV}{ {\rm keV} } \newcommand{\MeV}{ {\rm MeV} } \newcommand{\GeV}{ {\rm GeV} }

\newcommand{\sla}[1]{#1\!\!\!\!/ \,}

\newcommand{\lsp}{ \left ( } \newcommand{\rsp}{ \right ) } \newcommand{\Lg}{\mathcal{L}}  \newcommand{\To}{\Rightarrow}   

\newcommand{\vev}[1]{ \langle {#1} \rangle }

\newcommand{\Column}[3]{ \begin{pmatrix} #1 \\ #2 \\ #3 \end{pmatrix} }

\newcommand{\Diag}[3]{ \begin{pmatrix} #1 & 0 & 0 \\ 0 & #2 & 0 \\ 0 & 0 & #3 \\\end{pmatrix}}

\begin{document}

%%%%%%%%%%%%%%%%%%%%%%%%%%%%%%%%%%%%%%%%
\title{\Large Diagonal reflection symmetries, four-zero texture, and  \\
 trimaximal mixing with predicted $\theta_{13}$  in an $A_{4}$ symmeric model}
%and CP violation

\preprint{STUPP-21-245}
%%%%%%%%%%%%%%%%%%%%%%%%%%%%%%%%%%%%%%%%
\author{Masaki J. S. Yang}
\email{yang@krishna.th.phy.saitama-u.ac.jp}
\affiliation{Department of Physics, Saitama University, 
Shimo-okubo, Sakura-ku, Saitama, 338-8570, Japan}
%%%%%%%%%%%%%%%%%%%%%%%%%%%%%%%%%%%%%%%%

%\date{\today}

%%%%%%%%%%%%%%%%%%%%%%%%%%%%%
\begin{abstract} %%%%%%%%%%%%%%%%%%%%%
%%%%%%%%%%%%%%%%%%%%%%%%%%%%%

In this paper, we impose a magic symmetry on the neutrino mass matrix $m_{\nu}$ with universal four-zero texture and diagonal reflection symmetries. 
Due to the magic symmetry, the MNS matrix has trimaximal mixing inevitably.
Since the lepton sector has only six free parameters, physical observables of leptons are all determined from the charged leptons masses $m_{ei}$, the neutrino mass differences $\Delta m_{i1}^{2}$, and the mixing angle $\theta_{23}$.

This scheme predicts $\sin \th_{13} = 0.149$, that is almost equal to the latest best fit, as a function of the lepton masses $m_{e,\mu}$ and the mass differences $\Delta m_{i1}^{2}$. 
Moreover, even if the mass matrix has perturbations that break the magic symmetry, 
the prediction of $\sin\th_{13}$ is retained with good accuracy for the four-zero texture with diagonal reflection symmetries.

%%%%%%%%%%%%%%%%%%%%%%%%%%%%%
\end{abstract} %%%%%%%%%%%%%%%%%%%%%%
%%%%%%%%%%%%%%%%%%%%%%%%%%%%%

\maketitle

%%%%%%%%%%%%%%%%%%%%%%%%%%%%%%%%%%
\section{Introduction}
%%%%%%%%%%%%%%%%%%%%%%%%%%%%%%%%%%

To approach the flavor puzzle, a number of flavor structures have been considered.
In particular, universal texture \cite{Koide:2002cj,Koide:2003rx,Barranco:2010we,Zhou:2012ds} that imposes the same flavor structure on quark and leptons is quite appealing in the context of unified theories.
For this reason, various universal textures have been considered, such as four-zero texture $(M_{f})_{11} = (M_{f})_{13,31} = 0$  \cite{Fritzsch:1995nx, Chkareuli:1998sa, Nishiura:1999yt, Matsuda:1999yx, Fritzsch:1999ee, Chkareuli:2001dq, Fritzsch:2002ga, Xing:2003zd, Xing:2003yj, Bando:2004hi, Matsuda:2006xa, Branco:2006wv, Ahuja:2007vh, Xing:2015sva} and universal texture zero $(M_{f})_{11} = 0$ 
\cite{Albright:1989if,Rosner:1992qa,Roberts:2001zy,Grimus:2004hf,deMedeirosVarzielas:2018vab}.

Meanwhile, a set of generalized $CP$ symmetries (GCPs) \cite{Ecker:1980at, Ecker:1983hz, Gronau:1985sp, Ecker:1987qp,Neufeld:1987wa,Ferreira:2009wh,Feruglio:2012cw,Holthausen:2012dk,Ding:2013bpa,Girardi:2013sza,Nishi:2013jqa,Ding:2013hpa,Feruglio:2013hia,Chen:2014wxa,Ding:2014ora,Ding:2014hva,Chen:2014tpa,Chen:2015siy,Li:2015jxa,Turner:2015uta, Rodejohann:2017lre, Penedo:2017vtf,Nath:2018fvw, Yang:2020qsa} called diagonal reflection symmetries (DRS) has been proposed \cite{Yang:2020goc}.  These GCPs remove redundant $CP$ phases and enhance predictive power of flavor textures. 
By combining with the universal four-zero texture, the mixing angles and $CP$ phases of the CKM and MNS matrices are reproduced well with an accuracy of $10^{-3}$.
Since this system has eight parameters in both the quark and lepton sector, it can predict all physical quantities (such as Majorana phases $\a_{21}, \a_{31}$ and the mass of the lightest neutrino $m_{1}$) from current observables.
As a result, an approximate trimaximal mixing \cite{Harrison:2002kp,Harrison:2004he,Friedberg:2006it,Lam:2006wy,Bjorken:2005rm, He:2006qd, Grimus:2008tt,Li:2013jya,Channey:2018cfj,Verma:2019uiu,Bao:2021zwu} that is not assumed in the system is emerged. 
%. 
The trimaximal mixing is associated with a $Z_{2}$ symmetry called magic symmetry \cite{Lam:2006wy}. 
Therefore, in this paper, we investigate effects of imposing the magic symmetry on the four-zero texture with DRS. 
In addition, a field theoretical realization of such textures and perturbative breaking of the magic symmetry are discussed.

This paper is organized as follows. 
The next section gives a review of DRS and four-zero texture. 
In Sec.~3, we discuss a magic symmetry on the neutrino mass matrix, 
breaking of the symmetry, and its perturbative effects. 
In Sec.~4, the seesaw mechanism and a realization of magic symmetry are discussed. 
The final section is devoted to a summary. 

%%%%%%%%%%%%%%%%%%%%%%%%%%%%%%%%%%
\section{Diagonal reflection symmetries and universal four-zero texture}
%%%%%%%%%%%%%%%%%%%%%%%%%%%%%%%%%%

In this section, we review a previous study \cite{Yang:2020goc}. 
The following four-zero texture of mass matrices 
reproduce the mixing matrices and mass eigenvalues of fermions, 
\begin{align}
M_{u} = 
\begin{pmatrix}
0 & i \, C_{u} & 0 \\ 
- i \, C_{u} &\tilde B_{u} & B_{u} \\
0 & B_{u} & A_{u} 
\end{pmatrix} ,
~~~ 
m_{\n} = 
\begin{pmatrix}
0 & i \, c_{\n} & 0 \\ 
 i \, c_{\n} &\tilde b_{\n} & b_{\n} \\
0 & b_{\n} & a_{\n} 
\end{pmatrix} ,
~~~
M_{d,e}  =  
\begin{pmatrix}
0 & C_{d,e} & 0 \\ 
C_{d,e} &\tilde B_{d,e} & B_{d,e} \\
0 & B_{d,e}  & A_{d,e} 
\end{pmatrix} ,
\label{massmtrx}
\end{align}
with real parameters  $A_{f} \sim C_{f}$ and $a_{\n} \sim c_{\n}$. 
Hermiticity of Yukawa matrices is guaranteed by the parity symmetry in the left-right symmetric model \cite{Pati:1974yy,Senjanovic:1975rk,Mohapatra:1974hk}.
Eq.~(\ref{massmtrx}) also has the DRS \cite{Yang:2020goc}
\begin{align}
R \, M_{u}^{*} \, R = M_{u} , ~~~ R \, m_{\n}^{*} \, R = m_{\n}, ~~~
M_{d,e}^{*} = M_{d,e}, ~~~ R = {\rm diag} \, (-1,1,1) \, .
\label{refsym}
\end{align}
They are regarded as remnant symmetries such as the magic symmetry \cite{Lam:2006wy} of the neutrino mass matrix $m_{\n}$.  
For example, these symmetries are realized by vacuum expectation values (vevs) of two scalar fields with different phases $\vev{\th_{u}} = i v_{u}, ~ \vev{\th_{d}} = v_{d}$ coupled only in the first generation \cite{Yang:2020goc, Yang:2021smh}. 
A realization of DRS, magic symmetry, and zero textures is also discussed in Section 4 of this paper.  

The flavor mixing matrices are expressed by orthogonal matrices $O_{f}$ that diagonalize the mass matrices $M_{f}$ and $m_{\n}$ as follows, 
\begin{align}
V_{\rm CKM} = O_{u}^{T} \Diag{-i}{1}{1} O_{d}, ~~~
U_{\rm MNS} = O_{e}^{T} \Diag{+i}{1}{1} O_{\n}. 
\label{CKMMNS}
\end{align}
This system has eight parameters for both quarks and leptons. 
These mixing matrices (\ref{CKMMNS}) reproduce experiments with an accuracy of  $O(10^{-3})$. 
Although this system has an obvious deviation in $|V_{ub}|$, three-zero texture $(M_{u})_{11} \neq 0$ with DRS predicts CKM matrices with an  accuracy of $O(10^{-4})$ \cite{Yang:2021smh}.

As input parameters in the lepton sector, we choose the following eight observables; 
three charged lepton masses at mass of $Z$ boson $m_{Z}$ \cite{Xing:2011aa}, 
\begin{align}
m_{e} &= 486.570 \, [\keV],  ~~~
m_{\m} = 102.718 \, [\MeV], ~~~
m_{\t} = 1746.17 \, [\MeV], 
\end{align}
three mixing angles \cite{Esteban:2020cvm}, 
\begin{align}
\sin^{2} \th_{12} = 0.304^{+0.012}_{-0.012} , ~~~ 
\sin^{2} \th_{23} &= 0.573^{+0.016}_{-0.020} , ~~~
\sin^{2} \th_{13} = 0.02219^{+0.00062}_{-0.00063} , ~~~
%\d^{\rm MNS} = 197^{\circ} . 
\label{35}
\end{align}
and the mass-squared differences for the normal ordering \cite{Esteban:2020cvm}
\begin{align}
\D m_{21}^{2} &= 74.2^{+2.1}_{-2.0} \, [\meV^{2}], ~~~ 
\D m_{31}^{2} =  2517^{+26}_{-28} \, [\meV^{2}]. 
\label{massdiff}
\end{align}
The errors of neutrino parameters are in the range of 1 $\s$ region.
The inverted ordering is excluded because it is inconsistent with the four-zero texture.

A reconstructed neutrino mass matrix is \cite{Yang:2020goc, Yang:2021smh} 
\begin{align}
m_{\n}^{r} = 
\begin{pmatrix}
0 & 8.80 \, i & 0 \\
 8.80 \, i & 29.6 & 26.3 \\
0 & 26.3 & 14.1 
\end{pmatrix} , 
~~~ 
M_{e}^{r} \simeq 
\begin{pmatrix}
0 & \mp \sqrt{m_{e} m_{\m}} & 0 \\
\mp \sqrt{m_{e} m_{\m}} & \pm m_{\m} + t_{\t}^{2} \, m_{\t} & t_{\t} \, m_{\t} \\
0 & t_{\t} \, m_{\t} & m_{\t}
\end{pmatrix} , 
\label{Mn0}
\end{align}
where $t_{\t} \equiv \sin \t \simeq 0.06$ is a 23 mixing. 
The sign of 12 element ${\rm sign} (C_{e})$ is related to that of 22 element ${\rm sign} (B_{e}) = {\rm sign} (m_{e2})$ 
in order to keep the correct sign of the Jarlskog invariant $J_{\rm MNS}$  \cite{Jarlskog:1985ht}. 
From a viewpoint of unification, the other solution with $m_{1} \simeq 6.2$ [meV] is excluded because it predicts $(M_{e})_{22} \simeq m_{\t}$. 

An absolute value of the MNS matrix is calculated as
\begin{align}
|U_{\rm MNS}|
&= 
\begin{pmatrix}
0.8251 & 0.5449 & 0.1490 \\
0.2755 & 0.6031 & 0.7485 \\
0.4932 & 0.5825 & 0.6461 
\end{pmatrix}  
\label{UMNS2}  , 
\end{align}
with errors of about $O(10^{-3})$ from the best fit values. 
The predicted MNS matrix has an approximate trimaximal mixing, 
that is not assumed in the texture Eq.~(\ref{massmtrx}). Its cause is explored in the next section. 

%%%%%%%%%%%%%%%%%%%%%%%%%%%%%%%%%%
\section{Trimaximal mixing and magic symmetry}
%%%%%%%%%%%%%%%%%%%%%%%%%%%%%%%%%%

%
The matrix $m_{\n}^{r}$~(\ref{Mn0}) approximately has an eigenvector $\bm v \sim (1,-1, 1)$
and predicts the trimaximal mixing \cite{Harrison:2002kp, Bjorken:2005rm, He:2006qd}. 
The following symmetric matrix always predicts the trimaximal mixing \cite{Harrison:2004he, Friedberg:2006it, Lam:2006wy}
\begin{align}
m_{T} = 
\begin{pmatrix}
A & B & C  \\
B & D & A+C-D \\
C & A+C-D & B+D-C
\end{pmatrix} ,
\end{align}
with complex parameters $A \sim D$. 
A matrix will be called {\it magic} if the row sums and the column sums are all equal to a number $\a$ \cite{Lam:2006wy}. 
The matrix $m_{T}$ satisfies the following $Z_{2}$ symmetry 
\begin{align}
S_{2} \, m_{T} \, S_{2}^{T} = m_{T} , 
~~~ S_{2} 
=
\begin{pmatrix}
 \frac{1}{3} & -\frac{2}{3} & -\frac{2}{3} \\[2pt]
- \frac{2}{3} & \frac{1}{3} & -\frac{2}{3} \\[2pt]
- \frac{2}{3} & -\frac{2}{3} & \frac{1}{3} \\
\end{pmatrix} ,
~~ S_{2}^{2} = 1_{3}, 
\label{magic}
\end{align}
that is called {\it magic} symmetry \cite{Lam:2006wy}.

After a phase redefinition, 
the mass matrix $m_{\n}^{r}$~(\ref{Mn0}) can approximately be parameterized 
to a matrix with the (deformed) magic symmetry and zero textures \cite{Gautam:2016qyw},
\begin{align}
m_{\n T}  & \equiv
\begin{pmatrix}
0 & c & 0 \\
c & b+c & b+c \\
0 & b+c & b 
\end{pmatrix} \, .
\label{mnT}
\end{align}
Here, $b$ and $c$ are real parameters. A realization of the texture~(\ref{mnT}) has been discussed in a model with $A_{4}$ symmetry~\cite{Gautam:2016qyw}. 
The condition~(\ref{magic}) for the matrix $m_{\n T}$~(\ref{mnT})  is deformed to be 
\begin{align}
\tilde S_{2} \, m_{\n T} \, \tilde S_{2}^{T} = m_{\n T} , 
~~~\tilde S_{2} 
=
\begin{pmatrix}
 \frac{1}{3} & \frac{2}{3} & -\frac{2}{3} \\[2pt]
 \frac{2}{3} & \frac{1}{3} & \frac{2}{3} \\[2pt]
- \frac{2}{3} & \frac{2}{3} & \frac{1}{3} \\
\end{pmatrix} ,
~~ \tilde S_{2}^{2} = 1_{3}, 
\label{magic2}
\end{align}
%
% $(m_{\n T})_{i1} - (m_{\n T})_{i2} + (m_{\n T})_{i3} = a$ 
that is equivalent to the condition for an eigenvector 
\begin{align}
m_{\n T} \Column{1}{-1}{1} =  - c \Column{1}{-1}{1} .
\label{trimaxcond}
\end{align}
By a proper phase transformation, the generator 
$\tilde S_{2}$ in Eq.~(\ref{magic2}) and the DRS commute. 
Then, it composes a $Z_{2}$ symmetry only for the neutrinos.
Similar observation is found in 
$Z_{2} \times Z_{2}$ symmetry \cite{Lam:2006wm,Lam:2007qc,Lam:2008rs, Gupta:2011ct} and 
trimaximal $\m-\t$ reflection symmetry \cite{Rodejohann:2017lre}. 

The mass matrix $m_{\n T}$ is exactly diagonalized by so-called $TM_{2}$ mixing \cite{Albright:2008rp,Albright:2010ap}, 
which is a combination of  the tri-bi-maximal \cite{Harrison:2002er} and a 13 mixing
\begin{align}
O_{13}^{T} \, U_{\rm TBM}^{T} \, m_{\n T} \, U_{\rm TBM} \, O_{13}
= m_{\n T}^{\rm diag}, 
%= \Diag{b-\sqrt{a^2-a b+b^2}}{-a}{b+\sqrt{a^2-a b+b^2}} , 
\label{diag1}
\end{align}
where
\begin{align}
U_{\rm TBM} 
= 
\begin{pmatrix}
 \sqrt{\frac{2}{3}} & \frac{1}{\sqrt{3}} & 0 \\
 \frac{1}{\sqrt{6}} &-\frac{1}{\sqrt{3}} & \frac{1}{\sqrt{2}} \\
 -\frac{1}{\sqrt{6}} & \frac{1}{\sqrt{3}} & \frac{1}{\sqrt{2}} \\
\end{pmatrix} , 
~~~
O_{13} =
\begin{pmatrix}
\cos \ph_{13} & 0 &  \sin \ph_{13} \\
0 & 1 & 0 \\
- \sin \ph_{13} & 0 & \cos \ph_{13}
\end{pmatrix} , 
~~~
\tan 2 \ph_{13} = {\sqrt 3 \,  c \over 2 b + c } ,  
\label{diag2}
\end{align}
and 

\begin{align}
m_{\n T}^{\rm diag} = {\rm diag} (b+c -\sqrt{b^2 + b c+ c^2} \, , \,  -c \, , \, b+c+\sqrt{b^2 +  b c + c^2}) .  
\label{mndiag}
\end{align}
From Eq.~(\ref{mndiag}),  the two mass differences $\D m_{ij}^{2}$
are written by $b$ and $c$,
\begin{align}
\D m_{31}^{2} &= 4 (b+c) \sqrt{b^2+ b c +c^2} ,  \\
\D m_{21}^{2} &= (b+c) \, ( 2 \sqrt{b^2+b c +c^2}-2 b - c ) .
\end{align}
Conversely,  the parameters $b$ and $c$ are determined from 
the best fits of $\D m_{i1}^{2}$~(\ref{massdiff})  as
\begin{align}
(c, \, b) =
(9.19, \, 17.5) \, [\meV] , ~{\rm or}~  (-11.3, \, 33.0) \, [\meV] . 
\label{ab}
\end{align}
We exclude the second solution with $c<0$, 
because it corresponds to the solution $m_{1} = 6.2 \, [\meV]$ in Eq.~(\ref{Mn0}) 
that requires $(M_{e})_{22} \simeq m_{\t}$. 

A reconstructed mass matrix with phases is
\begin{align}
m_{\n T}^{r} & =
\begin{pmatrix}
 0 & 9.19 \, i & 0 \\
 9.19 \, i  & 26.7 & 26.7 \\
 0 & 26.7 & 17.5 \\
\end{pmatrix} \, [\meV]
=
\begin{pmatrix}
0 & i \, m_{2}  & 0 \\
i \, m_{2} & {m_{3} + m_{1} \over 2} & {m_{3} + m_{1} \over 2}  \\
0 & {m_{3} + m_{1} \over 2}  &{m_{3} + m_{1} \over 2} - m_{2}
\end{pmatrix} .
\label{mnT2}
\end{align}
Mass eigenvalues~(\ref{mndiag}) are found to be
\begin{align}
%(m_{\n 1}, \, m_{\n 2}, \, m_{\n 3}) 
m_{\n T}^{\rm diag} = {\rm diag} \, (3.21, \, 9.19, \, 50.3) \, [\meV]. 
\end{align}

The MNS matrix~(\ref{CKMMNS}) is approximately expressed as 
\begin{align}
U_{\rm MNS} &= O_{e}^{T} \Diag{+i}{1}{1} O_{\n} \\
& \simeq 
\begin{pmatrix}
1 & \mp  \sqrt{m_{e} / m_{\m}} & 0 \\
\pm \sqrt{m_{e} / m_{\m}} & 1 & 0 \\
0 & 0 & 1
\end{pmatrix} 
\begin{pmatrix}
+i & 0 & 0 \\
0 & c_{\t} & s_{\t} \\
0 & - s_{\t} & c_{\t} \\
\end{pmatrix}
U_{\rm TBM} \, O_{13} \Diag{\mp1}{1}{1} , 
\label{U30}
\end{align}
where $s_{\t} \equiv \sin \t , \, c_{\t} \equiv \cos \t$. 
The symbol $\pm$ denotes the sign of $C_{e}$~(\ref{Mn0}). 
To retain the correct sign of the Jarlskog invariant, 
a diagonal phase matrix diag$(\pm 1, 1,1)$ is added.  
By neglecting a small parameter $s_{\t}$, 
$U_{e3}$ is predicted as a function of the lepton masses $m_{e,\m}$ and the mass differences $\D m_{i1}^{2}$; 
\begin{align}
%| U_{\n 13} | = \sqrt{{1\over 3} -{2 b-a \over 6 \sqrt{a^2-a b+b^2}} } .
U_{e3} & \simeq 
\sqrt{\frac{m_e}{2 m_{\mu}}} c_{ \phi_{13}}  + \sqrt{\frac{m_e}{6 m_{\mu}}} s_{\phi_{13}}  + i  \sqrt{\frac{2}{3}} s_{\phi_{13}} , ~~~ |U_{e3} |  \simeq 0.149 ,  \label{Ue3} \\
%%%
\sin \ph_{13} & =  \sqrt{{1\over 2} -{2b + c \over 4 \sqrt{ b^2 + b c + c^{2}}} } \label{sinph13} \, . 
\end{align}
The last free parameter $s_{\t}$ is determined from 23 and 33 elements of $U_{\rm MNS}$ that have relatively good accuracy.
\begin{align}
|(U_{\rm MNS})_{23} / (U_{\rm MNS})_{33}| = \tan \th_{23}
 ~~ \To ~~ s_{\t} \simeq 0.0259 \, .
\end{align}
Therefore, $\th_{12}$ is predicted by taking $\th_{23}$ as an input parameter;
\begin{align}
\sin^{2} \th_{12}' = {|(U_{\rm MNS})_{12}|^{2} \over 1- |(U_{\rm MNS})_{13}|^{2}} =  0.341 \, , ~~ \sin \th_{12}' = 0.584 \, . 
\label{th12}
\end{align}
This is (barely) in the 3 $\s$ region of the best fit value~(\ref{35}). 
Reconstructed $U_{\rm MNS}$ is found to be 
\begin{align}
|U_{\rm MNS}| = 
\begin{pmatrix}
 0.803 & 0.577 & 0.149 \\
 0.300 & 0.592 & 0.748 \\
 0.516 & 0.562 & 0.646 \\
\end{pmatrix} .
\end{align}
All absolute values of components are within 3$\s$ range of global fit.  
Moreover, the value of $U_{e3}$ is very closer to the best fit value,  
$\sin 8.57^{\circ} = 0.1490$. 
Errors from the best fit are only about $O(10^{-2})$.
\begin{align}
|U_{\rm MNS}^{\rm best}| - |U_{\rm MNS}| =
\begin{pmatrix}
-0.022 & 0.033 & 0.000 \\
0.027 & -0.013 & 0.000 \\
0.021 & -0.018 & 0.000
\end{pmatrix} .
\end{align}
These errors can be improved by allowing small breaking of the trimaximal conditions Eq.~(\ref{magic2}) or (\ref{trimaxcond}).

%%%%%%%%%%%%%%%%%%%%%%%
\subsection{Breaking of the magic symmetry}
%%%%%%%%%%%%%%%%%%%%%%%

The four-zero texture with DRS and trimaximal condition predicts somewhat too large $\th_{12}'$~(\ref{th12}).
To fix this discrepancy, here we parameterize breakings of the magic symmetry.
A deviation of the reconstructed mass matrix~(\ref{mnT2}) from the best fit~(\ref{Mn0}) is
\begin{align}
\D m_{\n} = 
m_{\n}^{r} - m_{\n T}^{r} 
= -
\begin{pmatrix}
0 & 0.39 \, i  & 0 \\
0.39\, i & -2.9 & 0.04 \\
0 & 0.04 & 3.4
\end{pmatrix} .
\label{diff}
\end{align}
Thus, for $m_{\n T}$~(\ref{mnT}), we define breaking parameters $\d$ and $\e$ as follows;
\begin{align}
\d m_{\n} = \Diag{0}{\d}{\e}  , ~~~
m_{\n}' = m_{\n T} + \d m_{\n} = 
\begin{pmatrix}
0 & c & 0 \\
c & b + c + \d & b + c \\
0 & b + c & b+ \e
\end{pmatrix} \, . 
\label{dmn}
\end{align}
Since $m_{\n}'$ in Eq.~(\ref{dmn}) has four parameters, it describes four-zero texture with DRS~(\ref{massmtrx}) without loss of generality. 

Next, we will survey effects of perturbations $\d$ and $\e$ in the diagonalization of $m_{\n}'$. 
Defining a matrix $O = U_{\rm TBM} \, O_{13}$ in Eqs.~(\ref{diag1}) and (\ref{diag2}),
we obtain 
\begin{align}
O^{T} \, m_{\n T} \, O
= m_{\n T}^{\rm diag} . 
\label{m0}
\end{align}
Let $\d O$ and $\d m_\n^{\rm diag}$ be perturbative corrections to the orthogonal and eigenvalue matrices.
The diagonalization of the full mass matrix $m_{\n}'$ is written as
\begin{align}
(O^{T} + \d O^{T}) (m_{\n T} + \d m_\n) (O + \d O) = m_{\n T}^{\rm diag} + \d m_\n^{\rm diag} . 
\label{42}
\end{align}
By subtracting Eq.~(\ref{m0}) from Eq.~(\ref{42}), an expression for the first-order perturbation 
is found to be
\begin{align}
O^{T} m_{\n T} \d O + O^{T} \d m_\n \, O + \d O^{T} m_{\n T} \, O = \d m_{\n}^{\rm diag} .
\end{align}
The characteristic equation $m_{\n T} O = O m_\n^{\rm diag}$ from Eq.~(\ref{m0}) 
%
%\begin{align}
%m_\n^{\rm diag} O^{T} \d O + O^{T} \d m_\n O + \d O^{T} O m_\n^{\rm diag} = \d m_\n^{\rm diag}
%\end{align}
%
and the orthogonality relation $\d O^{T}\, O + O^{T}\d O = 0$ lead to
\begin{align}
m_{\n T}^{\rm diag} O^{T} \d O + O^{T} \d m_\n O - O^{T} \d O m_{\n T}^{\rm diag} = \d m_\n^{\rm diag} . 
\end{align}
For diagonal elements, it is rewritten as
\begin{align}
 (\d m_\n^{\rm diag})_{ii} =  (O^{T} \d m_\n O)_{ii} \, ,
\end{align}
and for off-diagonal elements, 
%
%\begin{align}
%m_\n^{\rm diag} O^{T} \d O + O^{T} \d m_\n O - O^{T} \d O m_\n^{\rm diag} = 0 , 
%\end{align}
%
\begin{align}
%& m_{i} (O^{T} \d O)_{ij} + (O^{T} \d m_\n O)_{ij} - (O^{T} \d O)_{ij} m_{j} = 0_{ij} \\
%& (m_{i} - m_{j}) (O^{T} \d O)_{ij}  = -  (O^{T} \d m_\n O)_{ij} \\
& (O^{T} \d O)_{ij}  = - { (O^{T} \d m_\n O)_{ij} \over m_{i} - m_{j} } \, . 
\label{otdo}
\end{align}
They are equivalent to the usual perturbative relations in quantum mechanics.
Since $O^{T} (O+\d O) = 1 + O^{T} \d O$ holds, Eq.~(\ref{otdo}) represents a perturbative rotation in the diagonalized basis.

In particular, a correction to the 13 element is
\begin{align}
& (\d O)_{13}  = - \sum_{i=1}^{2} O_{1i} { (O^{T} \d m_\n O)_{i3} \over m_{i} - m_{3} } \, .
\end{align}
By neglecting $m_{1}$ and $m_{2}$, this equation can be transformed as 
\begin{align}
(\d O)_{13} & \simeq {1  \over  m_{3} } \sum_{i=1}^{2} O_{1i} (O^{T} \d m_\n O)_{i3} \\
%& \simeq {1  \over  m_{3} } \sum_{i=1}^{3} 
%[ O_{1i} (O^{T} \d m_\n O)_{i3} - O_{13} (O^{T} \d m_\n O)_{33} ] \\
& = {1  \over  m_{3} } 
[ ( \d m_\n O)_{13} - O_{13} (O^{T} \d m_\n O)_{33} ] \, . 
\end{align}
Since the first term vanishes $(\d m_{\n})_{1i}=0_{i}$ from Eq. (\ref{dmn}), 
a correction to $\sin \th_{13}$ eventually becomes
\begin{align}
(\d O)_{13}  \simeq 
[ - O_{13} { (O^{T} \d m_\n O)_{33}   \over  m_{3} } ]
\simeq - \sin \th_{13} {\d + \e \over 2 \, m_{3}} .
\end{align}
We used $O\simeq U_{\rm TBM}$ in the last equality.
Since $\d \simeq - \e$ holds in the best fit~(\ref{diff}), 
the first-order perturbations for $O_{13}$ is very small. 
Thus, the prediction of $\sin \th_{13}$~(\ref{Ue3}) is approximately retained for the best fit of $m_{\n}$~(\ref{Mn0}) with $b=(m_{\n})_{23} - (m_{\n})_{12}$ and $c = (m_{\n})_{12}$.

For comparable values of $\d, \e \sim 3$ [meV] and $m_{3} \sim 50$ [meV], 
an error to $\sin \th_{13}$ is estimated as 
\begin{align}
|\d O_{13}| \simeq 0.0045,  ~~ 
{\d | \sin \th_{13}| \over \sin \th_{13}} \simeq 0.03 . 
\end{align}
Therefore, even if the four-zero texture with DRS has perturbations that break the magic symmetry, it predicts the correct $\sin\th_{13}$ with good accuracy.
This result comes from the fact that $m_{3}$ is the largest eigenvalue and $\sin \th_{13}$ is relatively small.

%%%%%%%%%%%%%%%%%%%%%%%%%%%%
\section{Type-I seesaw mechanism and realization of magic symmetry}
%%%%%%%%%%%%%%%%%%%%%%%%%%%%

In a model with the type-I seesaw mechanism \cite{Minkowski:1977sc,GellMann:1980v,Yanagida:1979as}, 
the following matrices $Y_{\n}$ and $\tilde m_{\n T}$ that have the same forms as Eq.~(\ref{massmtrx}) and (\ref{mnT2}) 
\begin{align}
Y_{\n} = 
\begin{pmatrix}
0 & i \, C_{\n} & 0 \\ 
- i \, C_{\n} &\tilde B_{\n} & B_{\n} \\
0 & B_{\n} & A_{\n} 
\end{pmatrix} , 
~~~ 
\tilde m_{\n T} = 
\begin{pmatrix}
0 & i \, c & 0 \\
i \, c & b + c & b + c \\
0 & b + c & b 
\end{pmatrix} \, ,
\label{52}
\end{align}
predict a four-zero texture for the Majorana mass matrix of right-handed neutrinos $M_{R}$ \cite{Nishiura:1999yt,Fritzsch:1999ee};
\begin{align}
M_{R} &= {v^{2} \over 2} Y_{\n}^{T} \tilde m_{\n T}^{-1} Y_{\n} \\
& = {v^{2} \over 2}
\begin{pmatrix}
 0 & -\frac{i \, C_{\nu }^2}{c} & 0 \\
 -\frac{i \, C_{\nu }^2}{c} & \frac{C_{\nu } \left(2 \tilde B_{\nu } -2 B_{\nu }+C_{\nu }\right)}{c}+\frac{\left(B_{\nu }-C_{\nu }\right){}^2}{b} & \frac{C_{\nu } \left(B_{\nu }-A_{\nu }\right)}{c}+\frac{A_{\nu } \left(B_{\nu }-C_{\nu }\right)}{b} \\
 0 & \frac{C_{\nu } \left(B_{\nu }-A_{\nu }\right)}{c}+\frac{A_{\nu } \left(B_{\nu }-C_{\nu }\right)}{b} & \frac{A_{\nu }^2}{b} \\
 \end{pmatrix} .
\end{align}
Moreover, the matrix $M_{R}$ also has the DRS~(\ref{refsym}), $R \, M_{R}^{*} \, R = M_{R}$. 
A hierarchical $Y_{\n}$ with $A_{\n} \gg B_{\n} , \tilde B_{\n} \gg C_{\n}$ yields 
a strongly hierarchical $M_{R}$. 
\begin{align}
M_{R} \sim {v^{2} \over 2}
\begin{pmatrix}
 0 & i \frac{C_{\nu }^2}{c} & 0 \\
i \frac{C_{\nu }^2}{c} & \frac{B_{\nu }^2}{b} & \frac{A_{\nu } B_{\nu }}{b} \\
 0 & \frac{A_{\nu } B_{\nu }}{b} & \frac{A_{\nu }^2}{b} \\
\end{pmatrix} .
\label{36}
\end{align}
However,  from Eq.~(\ref{36}), it seems difficult to derive a condition for the magic symmetry.

The mass matrix $\tilde m_{\n T}$~(\ref{52}) has the following texture and symmetry, 
\begin{itemize}
\item {\it four-zero texture,} 
\item {\it diagonal reflection symmetry,}
\item {\it (deformed) magic symmetry,}
\end{itemize}
that are retained in the type-I seesaw mechanism. 
By imposing all these conditions on the Yukawa matrix $Y_{\n}$, 
the matrix $M_{R}$ exhibits the same properties and is written by two parameters,
\begin{align}
Y_{\n} = 
\begin{pmatrix}
0 & i \, C_{\n} & 0 \\
- i \, C_{\n} & C_{\n} + B_{\n} & C_{\n} + B_{\n} \\
0 & C_{\n} + B_{\n} & B_{\n} 
\end{pmatrix}  , 
 ~~~
M_{R}  =  
\begin{pmatrix}
0 & - i \, C_{R} & 0 \\
- i \, C_{R} & C_{R}+ B_{R} & C_{R}+B_{R} \\
0 & C_{R} + B_{R} & B_{R} 
\end{pmatrix}  .
\label{magicYM}
\end{align}
In this case, the light neutrino mass $\tilde m_{\n}$ is obtained as 
\begin{align}
\tilde m_{\n} &= {v^{2} \over 2} Y_{\n} M_{R}^{-1} Y_{\n}^{T} \\
&= {v^{2} \over 2}
\begin{pmatrix} 
 0 & i {C_{\n}^2 \over C_{R}} & 0 \\
i {C_{\n}^2 \over C_{R}} & {C_{\n}^2 \over C_{R}} + {B_{\n}^2 \over B_{R} } & 
{C_{\n}^2 \over C_{R}} + {B_{\n}^2 \over B_{R} } \\
 0 &{C_{\n}^2 \over C_{R}} + {B_{\n}^2 \over B_{R} } &  {B_{\n}^2 \over B_{R} } \\
\end{pmatrix} ,
\end{align}
with $v = 246 [\GeV]$. 
Indeed this $\tilde m_{\n}$ is magic, diagonal reflection symmetric, and has four-zero texture. 
The parameters of $M_{R}$ are concisely expressed by that of $\tilde m_{\n}$ and $Y_{\n}$ as
\begin{align}
B_{R} = {v^{2} \over 2} {B_{\n}^2 \over b } , ~~~ 
C_{R} = {v^{2} \over 2} {C_{\n}^{2} \over c } . 
\end{align}
Therefore, in this scheme, the neutrino sector ($\tilde m_{\n}, Y_{\n},$ and $M_{R}$) has only two free parameters. 

Diagonalization of $Y_{\n}$ and $M_{R}$ have the same form to that of $m_{\n T}$,~Eq.~(\ref{diag1}) and (\ref{diag2}). 
Therefore, mass eigenvalues of $M_{R}$ can be written in the same way as Eq.~(\ref{mndiag}), 
\begin{align}
M_{R}^{\rm diag} = {\rm diag} ( B_{R} + C_{R} -\sqrt{B_{R}^2 + B_{R} C_{R}+ C_{R}^2} \, , \,  - C_{R} \, , \, B_{R} + C_{R} + \sqrt{B_{R}^2 + B_{R} C_{R} + C_{R}^2}) .  
\end{align}
%
%Unfortunately, general analyses with  perturbations $\d m_{\n}$ become more complicated. 
%we leave it our future work.

%%%%%%%%%%%%%%%%%%%%%%
\subsection{Realization of symmetries and texture}
%%%%%%%%%%%%%%%%%%%%%%

In order to justify the symmetries assumed above, we consider a partial compositeness-like realization by the two Higgs doublet model (2HDM) with $A_{4}$ flavor symmetry. 
The $A_{4}$ flavor symmetry \cite{Ma:2002yp,Altarelli:2004za} has been studied in a wide range of models \cite{Ma:2002yp,Altarelli:2004za,Chen:2005jm,Ma:2006wm,Ma:2006sk,Ma:2005qf,Altarelli:2005yx,Adhikary:2006wi,Ma:2006vq,Lavoura:2006hb,King:2006np,Lavoura:2007dw,Honda:2008rs,Brahmachari:2008fn,Bazzocchi:2008sp,Bazzocchi:2007na,Bazzocchi:2007au,Adhikary:2008au,Hagedorn:2009jy,Ciafaloni:2009ub,Branco:2009by,Bazzocchi:2008rz,Feruglio:2009hu,delAguila:2010vg,Barry:2010zk,Altarelli:2010gt,Ishimori:2012fg,GonzalezFelipe:2013xok,Ferreira:2013oga,Ma:2015pma,Petcov:2017ggy}. 
The group consists of the following generators $S$ and $T$;
\begin{align}
 S^{2} = T^{3} = (ST)^{3} = 1 \, . 
\end{align}
There are four irreducible representations $1, 1', 1''$ and $3$. For the $3$ dimensional   representation, $S$ and $T$ are usually taken as follows
\begin{align}
S = \Diag{1}{-1}{-1}, ~~~ 
T = 
\begin{pmatrix}
0 & 1 & 0 \\
1 & 0 & 0 \\
0 & 0 & 1
\end{pmatrix} \, . 
\label{ST}
\end{align}

The basic idea of the partial compositeness \cite{Kaplan:1991dc, Contino:2006nn, Agashe:2004rs, Contino:2006qr, Agashe:2008fe, Contino:2010rs, KerenZur:2012fr, Redi:2013pga} is that the SM fields at low energy are the mixed states between elemental (massless) fields and composite (massive) fields, like $\rho - \g$ mixing. 
Flavor structures are induced from mixings between massive and massless fermions with the same quantum numbers. 

Let us consider a model with $A_{4}$ flavor symmetry and field contents as in Table 1.
$U(1)_{PQ}$ is a chiral symmetry that distinguishes several particle species, and GCP is a  generalized CP symmetry.
Since the GCP charge of a field with nontrivial transformations under $A_{4}$ is $1$, we do not need to consider the consistency conditions \cite{Feruglio:2012cw, Holthausen:2012dk}.
$H$ and $H_{2}$ are two Higgs doublets with different chiral charges,  
$L, E, L', N', E'$ are fields with  heavy masses and the same charges as the corresponding SM fields.
$\varphi, \varphi'$, and $\D$ are flavons with $3$ and $1'$ representation, $\Phi$ is a scalar field that breaks $U(1)_{PQ}$.

\begin{table}[h]
  \begin{center}
    \begin{tabular}{|c|ccccc|} \hline
           & $SU(2)_{L}$ & $U(1)_{Y}$ & $A_{4}$ & $U(1)_{PQ} $ & GCP \\ \hline \hline
      $l_{Li}$ & \bf 2 & $-1/2$ & 3& 0  & 1\\
      $\n_{Ri}$ & \bf 1 & $0$ & 3 & 0  & 1\\ 
      $e_{Ri}$ & \bf 1 & $-1$ & 3 &$-2$  &  1 \\ 
      $H$ & \bf 2 & $1/2$ & 1 & $0$ & 1  \\ 
      $H_{2}$ & \bf 2 & $1/2$ & 1 & $2$ & $- 1$\\ \hline  
      $L_{(L,R)}$ & \bf 2 & $-1/2$ & 1 & $-1$ & 1 \\
      $E_{(L,R)}$ & \bf 1 & $-1$ & 1 & $-1$ & 1 \\
      $N'_{(L,R)}$ & \bf 1 & 0 & $1'$ & $0$ & 1 \\  
      $L'_{L}$ & \bf 2 & $-1/2$ & $1'$  & $0$ & 1 \\
      $L'_{R}$ & \bf 2 & $-1/2$ & $1''$  & $0$ & 1 \\
      $E'_{L}$ & \bf 1 & $ -1 $ & $1'$ & $- 2$   & 1 \\ 
      $E'_{R}$ & \bf 1 & $ -1 $ & $1'$ & $0$   & 1 \\ \hline
      $\varphi$ & \bf 1 & $1$ & 3 & $1$ & 1 \\ 
      $\varphi'$ & \bf 1 & $1$ & 3 & $0$ & 1\\ 
      $\Delta$ & \bf 1 & $1$ & $1'$ & $0$ & 1\\ 
      $\Phi$ & \bf 1 & $1$ & 1 & $-2$ & 1\\ \hline
%      $\tilde \varphi'$ & \bf 1 & $1$ & 3 & $2$ & 1  \\  \hline
    \end{tabular}
    \caption{Charge assignments of fields under gauge, flavor, and GCP symmetries. }
  \end{center}
\end{table}

Similar to the simplified two-site description of composite Higgs model \cite{Contino:2006nn}, 
the most general Lagrangian under the symmetry imposed on the model 
 is divided into three parts: 
\begin{align}
\Lg_{\rm heavy}& = 
\bar{L} (i \sla{D} - M_{L}) L  + \bar{E} (i \sla{ D} - M_{E}) E  +  \bar{N'} (i \sla{ D} - M_{N'}) N' + \bar{L'} i \sla{D}  L'   + \bar{E'} i \sla{ D} E'  \\
%%%
& - ( Y^{E} \bar L_{L} H E_{R} + Y^{N'} \bar L'_{L} \widetilde{H} N'_{R} + Y^{E'} \bar L'_{L} H E'_{R} + \widetilde{Y}^E \bar{E}_{L} \widetilde H L_{R} + {\rm h.c.} ) \label{compYukawa} \\
% - \widetilde{Y}^{N'} \bar{N}'_{L} H L'_{R} - \widetilde{Y}^{E'} \bar{E}'_{L} \widetilde H L'_{R} ) \label{compYukawa}
& -  \lsp 
 y_{L'} \bar L_{L}' L_{R}' \D^{*} + y_{E'} \bar E_{L}' E_{R}' \Phi  
+ {y_{NL} \over 2} \D^{*} \bar N'_{L} N_{L}^{\prime \, c} 
+ {y_{NR} \over 2} \D \bar N_{R}^{\prime \, c} N'_{R} 
+{\rm h.c.} \, \rsp \label{compYukawa2}
\, ,  \\
\Lg_{\rm light} & = 
\bar l_{Li}  i  \sla D l_{Li} + \bar \n_{Ri}  i  \sla D \n_{Ri} + \bar e_{Ri}  i  \sla D e_{Ri} \,   \\
 & - y_{e 0} \bar l_{L i} e_{R i} H_{2} - y_{\n 0} \bar l_{Li} \n_{Ri} \widetilde H  
- {1\over 2} m_{R 0 } \bar \n_{R i}^{c} \n_{Ri}  \, , 
\\
%%%
\Lg_{\rm mixing} &=  \Big (
\l^{L} \bar{l}_{Li} L_{R} \varphi_{i} + \l^{E} \bar{E}_{L} e_{Ri} \varphi_{i} 
%+ \l^{\n}_{ij} \bar{N}_{Li} \n_{Rj} + \l^{e}_{ij} \bar{E}_{Li} e_{Rj} 
 \\
& + \l^{L'} \bar{l}_{Li} L'_{R} P_{ij}^{*} \varphi'_{j} + \l^{N'} \bar{N}'_{L} \n_{Ri} P_{ij}^{*} \varphi'_{j}
+ \l^{E'} \bar{E}'_{L} e_{R i} P_{ij}^{*} \varphi'_{j} \Big)
+{\rm h.c.} \, , 
\end{align}
where $\tilde H \equiv i \s^{2} H^{*}$ is the conjugate field of the Higgs doublet $H$
and 
\begin{align}
P \equiv \Diag{1}{\o}{\o^{2}} \, . 
\end{align}
A combination such as $\n_{R i } P_{ij}^{*}\varphi_{j}'$ constitutes a $1'$ representation.

The scalar fields are assumed to have real vevs
\begin{align}
\vev{\varphi} = V_{\varphi} \Column{1}{1}{1} , ~~~
\vev{\varphi'} = V_{\varphi'} \Column{0}{1}{-1}  , ~~~ 
\vev{\Delta} = V_{\D} , ~~~
\vev{\Phi} = V_{\Phi}  , ~~~ 
\vev{H} = V_{H}, ~~~ 
\vev{H_{2}} = V_{H_{2}} , 
\end{align}
where $V_{X} \in \mathbb R$ for $X = \varphi, \varphi', \Delta, \Phi, H$ and $H_{2}$. 
The vevs of $\varphi, \varphi', \Delta,$ and $\Phi$ break $A_{4}$ and $U(1)_{PQ}$, 
but retain GCP symmetry.
For the Higgs doublets $H$ and $H_{2}$, the reality of vevs are 
achieved over a wide range of parameters. 
According to the spirit of partial compositeness, 
the masses of heavy fields are assumed to be larger than those of flavon's vevs, $M_{F} , \vev{\D}, \vev{\Phi} \gg \l_{F} \vev{\varphi}$. 
Due to this, the linear mixing terms $\Lg_{\rm mixing}$ induce mass terms between massive  and massless fields.

When the heavy fields are integrated out, 
the SM interactions at low energy are represented by seesaw-like formulae 
\footnote{
In order not to change the eigenstates of $N_{L,R}'$ significantly, the lepton number violating (LNV) parameters $y_{NL} \vev{\D}$ and $y_{NR} \vev{\D}$ are assumed as $M_{N'} \gg y_{NL} \vev{\D}, y_{NR} \vev{\D}$ in Eq.~(\ref{SMy3}). 
However, even if this does not hold and both $y_{NL}$ and $y_{NR}$ contribute to LNV, the final result remains the same because the flavor structure is only  generated by $\vev{\varphi}$ and $\vev{\varphi'}$.
}
\begin{align}
y_{e} &= y_{e0} 1_{3} r_{e} + \vev{\varphi} \l^{L} M_{L}^{-1} Y^{E} M_{E}^{-1} \l^{E} \vev{\varphi}^{T} + \vev{\varphi'} P^{*} \l^{L'} (y_{L'} \vev{\D})^{-1} Y^{E'} (y_{E'} \vev{\Phi})^{-1} \l^{E'} P^{*} \vev{\tilde \varphi'}^{T} , \label{SMy1}  \\
y_{\n} &= y_{\n 0} 1_{3} + \vev{\varphi'} P^{*} \l^{L'} (y_{L'} \vev{\D})^{-1} Y^{N'} M_{N'}^{-1} \l^{N'} P^{*} \vev{\varphi'}^{T},  \\
m_{R} &= m_{R 0} 1_{3} + 
\vev{\varphi'} P^{*} (\l^{N'})^{*} (y_{NL} \vev{\D})^{-1} \l^{N'} P^{*}  \vev{\varphi'}^{T} \, . 
\label{SMy3}
\end{align}
Here, $r_{e} = \vev{H_{2}} / \vev{H}$ is a factor that takes into account the ratio of the vevs of the two Higgs doublets.
The term with $\widetilde Y^{E}$ in Eq.~(\ref{compYukawa}) does not contribute SM matrices in the first order of $\l^{f} \vev{\varphi} / M_{F}$. 
Figure 1 shows diagramatic explanations of Eq.~(\ref{SMy1}).
\begin{figure}[h]
\begin{center}
   \includegraphics[width=17cm]{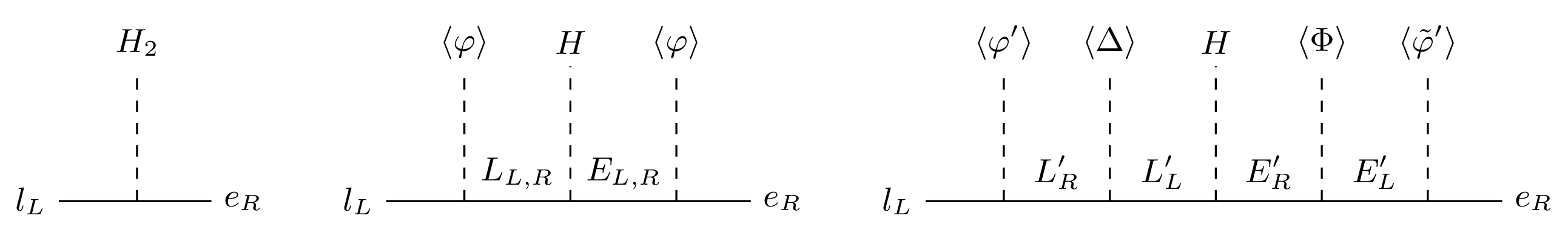} 
\caption{ A diagrammatic description of processes that generate the flavor structure of $y_{e}$.}
\label{fig1}
\end{center}
\end{figure}

The flavor structure by terms with flavons are
\begin{align}
\vev{\varphi} \otimes \vev{\varphi}^{T}  = 
\begin{pmatrix}
1 & 1 & 1 \\
1 & 1 & 1 \\
1 & 1 & 1 \\
\end{pmatrix} , 
~~~
\Diag{1}{\o^{2}}{\o}
\vev{\varphi'} \otimes \vev{\varphi'}^{T} 
\Diag{1}{\o^{2}}{\o} = 
\begin{pmatrix}
0 & 0 & 0 \\
0 & \o & -1 \\
0 & -1 & \o^{2}
\end{pmatrix} . 
\end{align}
The first term is the democratic matrix, that have been well discussed for a long time 
\cite{Harari:1978yi, Koide:1983qe, Koide:1989zt, Tanimoto:1989qh, Fritzsch:1989qm, Lehmann:1995br, Fukugita:1998vn, Tanimoto:1999pj, Haba:2000rf, Hamaguchi:2002vi, Kakizaki:2003fc, Kobayashi:2004ha, Fritzsch:2004xc, Jora:2006dh, Mondragon:2007af, Xing:2010iu, Zhou:2011nu, Canales:2012dr, Yang:2016esx, Yang:2016crz}. 

Combining all the flavor-independent coefficients, we obtain a complex symmetric matrix for Yukawa of leptons;
\begin{align}
y_{e} =
a_{e} 
\begin{pmatrix}
1 & 1 & 1 \\
1 & 1 & 1 \\
1 & 1 & 1 \\
\end{pmatrix} 
+ 
b_{e} 
\begin{pmatrix}
0 & 0 & 0 \\
0 & \o & -1 \\
0 & -1 & \o^{2}
\end{pmatrix} 
+ c_{e} \Diag{1}{1}{1} . 
\end{align}
On the other hand, since there is no contribution of the democratic matrix to the neutrinos, 
\begin{align}
y_{\n}, m_{R} =
b_{\n, R}' 
\begin{pmatrix}
0 & 0 & 0 \\
0 & \o & -1 \\
0 & -1 & \o^{2}
\end{pmatrix} 
+ c_{\n, R}' \Diag{1}{1}{1}  .
\end{align}
By the following basis transformation, 
these three mass matrices are converted to the four-zero texture 
\begin{align}
U^{T} y_{e} U= 
\begin{pmatrix}
0 & c_{e} & 0 \\
c_{e} & -b_{e} &  b_{e} \\
0 &  b_{e} & 3 a_{e} - b_{e} + c_{e}
\end{pmatrix}
, ~~ 
U^{T} (y_{\n} , m_{R}) U = 
\begin{pmatrix}
0 & c'_{\n , R} & 0 \\
c'_{\n , R} &- b'_{\n , R} &  b'_{\n, R} \\
0 & b'_{\n , R} &  - b'_{\n, R} + c'_{\n, R}
\end{pmatrix} , 
\label{UyU}
\end{align}
where
\begin{align}
U = {1 \over \sqrt 3}
\begin{pmatrix}
1 & 1 & 1 \\
\o & \o^{2} & 1 \\
\o^{2} & \o & 1
\end{pmatrix}  \, . 
\end{align} 

For $y_{\n}$ and $M_{R}$, a $Z_{2}$ symmetry due to $S$ in Eq.~(\ref{ST}) remains unbroken.
\begin{align}
 S (y_{\n}, m_{R}) S = (y_{\n} , m_{R}) \, . 
\end{align}
This $Z_{2}$ is changed to magic symmetry~(\ref{magic}) by the basis transformation. 
\begin{align}
- U^{\dg} S U =  
U^{\dg} \Diag{-1}{1}{1} U = 
\begin{pmatrix}
 \frac{1}{3} & -\frac{2}{3} & -\frac{2}{3} \\[2pt]
- \frac{2}{3} & \frac{1}{3} & -\frac{2}{3} \\[2pt]
- \frac{2}{3} & -\frac{2}{3} & \frac{1}{3} \\
\end{pmatrix}
= S_{2} \, . 
\end{align}

Finally, since a nontrivial GCP charge is imposed only on $H_{2}$ in Table 1, 
the GCP invariance restricts complex phases of couplings as 
\begin{align}
 c_{e}^{*} = (-1) c_{e},  ~~~ a_{e}^{*} = a_{e},   ~~~ b_{e}^{*} = b_{e},  ~~~ b_{f}^{*} = b_{f},  ~~~ c_{f}^{*} = c_{f}, ~~~ 
{\rm for} ~~ f = \n, R \, . 
\end{align}
Among these parameters, only $c_{e}$ is purely imaginary, 
and all other $a_{e}, b_{e}, b_{\n}, c_{\n}, b_{R}, c_{R}$ are real.  
In this basis, the GCP is broken by the real vev $\vev{H_{2}}$. 
By redefining the phases and parameters, 
Eq.~(\ref{UyU}) results in the four-zero texture with DRS and magic symmetry (\ref{massmtrx}) and (\ref{magicYM}). 
The hierarchy of lepton masses requires that a tiny value of $c_{e}$. 
The term with $c_{e}$ is forbidden by the $U(1)_{PQ}$ symmetry if $H_{2}$ does not exist.
Thus, for example, $c_{e}$ could be made naturally small by using heavy 3-representation fermions instead of $H_{2}$.

%%%%%%%%%%%%%%
\section{Summary}
%%%%%%%%%%%%%%

In this paper, we impose a magic symmetry on the neutrino mass matrix $m_{\nu}$ with universal four-zero texture and diagonal reflection symmetries. 
Due to the magic symmetry, the MNS matrix has inevitably trimaximal mixing.
Since free parameters are reduced by two by fixing the eigenvector, 
the lepton sector has only six free parameters.
Therefore, physical observables of leptons are all determined from the charged leptons masses $m_{ei}$, the neutrino mass differences $\Delta m_{i1}^{2}$, and the mixing angle $\th_{23}$.

This scheme predicts $\sin \th_{13} = 0.149$, that is almost equal to the latest best fit, as a function of the lepton masses $m_{e,\m}$ and the mass differences $\D m_{i1}^{2}$. 
Moreover, even if the mass matrix has perturbations that break the magic symmetry, 
this prediction of $\sin\th_{13}$ is retained with good accuracy for the four-zero texture with DRS. 

The diagonal reflection symmetries, four-zero texture, and magic symmetry are all seesaw-invariant. 
Therefore, the imposition of these conditions on the neutrino Yukawa matrix $Y_{\n}$ leads to the same structure for the mass of right-handed neutrinos $M_{R}$.
In this case, the neutrino sector has only two parameters, and $m_{\n}$ is concisely represented from  parameters of $Y_{\n}$ and $M_{R}$ by the type-I seesaw mechanism.

For the justification of the assumed symmetries, we considered a partial compositeness-like realization by the two Higgs doublet model with $A_{4}$ flavor symmetry.
By a vev of a 3 representation flavon $\varphi \propto (1,1,1)$, 
the democratic texture emerges only in the electron-type Yukawa matrix. 
The four-zero texture appears by proper basis transformation. 
Since a contribution of democratic texture does not exist in the neutrino sector, a $Z_{2}$ subgroup of $A_{4}$ is preserved. This remnant symmetry is identified with the magic symmetry by the basis transformation.

%%%%%%%%%%%%%%
\section*{Acknowledgement}
%%%%%%%%%%%%%%

This study is financially supported %by JSPS KAKENHI Grants
by JSPS Grants-in-Aid for Scientific Research
No.~18H01210, No. 20K14459,  
and MEXT KAKENHI Grant No.~18H05543.

%\bibliographystyle{bib/h-physrev50}
%\bibliography{bib/fourzero,bib/onezero,bib/trimaximal,bib/refsym,bib/GCP,bib/mutausym,bib/LR,bib/PSGUT,bib/U(2), bib/SU(3), bib/CompositeHiggs, bib/A4, bib/democratic}

\end{document}